\documentclass[12pt]{article}
\usepackage{epsfig}
\usepackage{pstricks}
\usepackage{amstext}
\usepackage{amsmath}
\usepackage{amssymb}
\begin{document}
\title{
{\bf A Lanczos approach to the inverse square root of a
large and sparse matrix}
}

\author{Artan Bori\c{c}i \\
        {\normalsize\it Paul Scherrer Institute}\\
        {\normalsize\it CH-5232 Villigen PSI}\\
        {\normalsize\it Artan.Borici@psi.ch}\\
}

\date{}
\maketitle

\begin{abstract}
I construct a Lanczos process on a large and sparse matrix and use
the results of this iteration to compute the inverse square root
of the same matrix.
The algorithm is a stable version of an earlier proposal by the author.
It can be used for problems related to the matrix sign and polar decomposition.
The application here comes from the theory
of chiral fermions on the lattice.
\end{abstract}

\section{Introduction}

The computation of the inverse square root of a matrix is a special problem
in scientific computing. It is related to the matrix sign and polar decomposition
\cite{Higham}.

One may define the {\it matrix sign} by:
\begin{equation}
\text{sign}(A) = A(A^2)  ^{-\frac{1}{2}}
\end{equation}
where $A$ is a complex matrix with no pure imaginary eigenvalues.

In polar coordinates, a complex number $z = x + i~y$, is represented by
\begin{equation}
z = |z|~e^{i~\phi}, ~~~\phi = \arctan \frac{y}{x}
\end{equation}
In analogy, the polar decomposition of a matrix $A$ is defined by:
\begin{equation}
A = V(A^{\dag}A)^{\frac{1}{2}}, ~~~V^{-1} = V^{\dag}
\end{equation}
where $V$ is the polar part and the second factor corresponds to the absolute
value of $A$.

The mathematical literature invloving the matrix sign function traces back to 1971
when it was used to solve the Lyapunov and algebraic Riccati equations \cite{Higham}.

In computational physics one may face a similar problem when dealing with Monte
Carlo simulations of fermion systems, the so-called {\it sign problem}
\cite{sign_problem}. In this case the integration measure is proportional to
the determinant of a matrix and the polar decomposition may be helpful to monitor
the sign of the determinant.

The example brought in this paper comes from the recent progress in formulating
Quantum Chromodynamics (QCD) on a lattice with exact chiral symmetry
\cite{Neuberger_review99}.

In continuum, the massless Dirac propagator $D_{cont}$ is chirally symmetric, i.e.
\begin{equation}
\gamma_5 D_{cont} + D_{cont} \gamma_5 = 0
\end{equation}

On a regular lattice with spacing $a$ the symmetry is suppressed
according to the Ginsparg-Wilson relation:
\cite{Ginsparg_Wilson}:
\begin{equation}
\gamma_5 D + D \gamma_5 = a D \gamma_5 D
\end{equation}
where $D$ is the lattice Dirac operator.

An explicit example of a Dirac operator obeying this relation
is the so-called overlap operator \cite{Neuberger1}:
\begin{equation}
a D = 1 - A (A^{\dag}A)^{-1/2}, ~~~~A = M - aD_{W}
\end{equation}
where $M$ is a shift parameter in the range $(0,2)$,
which I have fixed at one.

$D_{W}$ is the Wilson operator,
\begin{equation}
D_{W} = \sum_{\mu = 1}^4 \gamma_{\mu} \nabla_{\mu}
      - \frac{a}{2} \sum_{\mu = 1}^4 \Delta_{\mu}
\end{equation}
which is a nearest-neighbors discretization of the continuum Dirac operator
(it violates the Ginsparg-Wilson relation).
$\nabla_{\mu}$ and $\Delta_{\mu}$ are first and second
order covariant differences given by:
\begin{flushleft}
\hspace{4cm} $(\nabla_{\mu} \psi)_i = \frac{1}{2 a}
  (U_{\mu,i} \psi_{i+\hat{\mu}} - U_{\mu,i-\hat{\mu}}^{\dag} \psi_{i-\hat{\mu}})$

\hspace{4cm} $(\Delta_{\mu} \psi)_i = \frac{1}{a^2}
  (U_{\mu,i} \psi_{i+\hat{\mu}} + U_{\mu,i-\hat{\mu}}^{\dag} \psi_{i-\hat{\mu}} - 2 \psi_i)$
\end{flushleft}
where $\psi_i$ is a fermion field at the lattice site $i$ and $U_{\mu,i}$
an $SU(3)$ lattice gauge filed associated with the oriented link $(i,i+\hat{\mu})$.
These are unitary 3 by 3 complex matrices with determinant one.
A set of such matrices forms a lattice gauge ``configuration''.

$\gamma_{\mu}, \mu = 1, \ldots, 5$ are $4 \times 4$ Dirac matrices
which anticommute with each-other.

Therefore, if there are $N$ lattice points in four dimensions,
the matrix $A$ is of order
$12N$. A restive symmetry of the matrix $A$ that comes from the
continuum is the so called $\gamma_5-symmetry$ which is the Hermiticity
of the $\gamma_5 A$ operator.

By definition, computation of $D$ involves the inverse
square root of a matrix. This is a non-trivial task for large matrices.
Therefore several algorithms have been proposed by lattice QCD physicists
\cite{Bunk,Hern_Jans_Luesch,Neuberger2,Edwards_Heller_Narayanan,Borici}.

All these methods rely on matrix-vector multiplications
with the sparse Wilson matrix $D_{W}$,
being therefore feasible for large simulations.

In fact, methods that approximate the inverse square root by
Legendre \cite{Bunk} and Chebyshev polynomials \cite{Hern_Jans_Luesch}
need to know {\it apriori} the extreme eigenvalues of $A^{\dag}A$
to be effective. This requires computational resources of at least one
Conjugate Gradients (CG) inversion.

In \cite{Neuberger2} the inverse square root is approximated by a rational
approximation, which allows an efficient
computation via a multi-shift CG iteration.
Storage here may be an obstacle which is remedied by a second CG step
\cite{Neuberger3}.

The Pade approximation used by \cite{Edwards_Heller_Narayanan} needs
the knowledge of the smallest eigenvalue of $A^{\dag}A$. Therefore the method becomes
effective only in connection with the $D$ inversion \cite{Edwards_Heller_Narayanan_2}.

The method presented earlier by the author \cite{Borici} relies on taking
exactly the inverse square root from the Ritz values. These are the roots of the
Lanczos polynomial approximating the inverse of $A^{\dag}A$.

In that work the Lanczos polynomial was constructed by applying the Hermitian
operator $\gamma_5 A$. The latter is indefinite, thereby responsible for
observed oscillations in the residual vector norm \cite{Borici}.

Here I use a Lanczos polynomial on the positive definite
matrix $A^{\dag}A$. In this case the residual
vector norm decreases monotonically and leads to a stable method.
This is a crucial property that allows a reliable stopping criterion that
I will present here.

The paper is selfcontained: in the next section I will briefly
present the Lanczos algorithm and set the notations.
In section 3, I use the algorithm to solve linear systems,
and in section 4, the computation of the inverse square root is given.
The method is tested in section 5 and conclusions are drawn in the end.

\section{The Lanczos Algorithm}

The Lanczos iteration is known to approximate the spectrum of
the underlying matrix in an optimal way and, in particular,
it can be used to solve linear systems \cite{Golub_VanLoan}.

Let $Q_n = [q_1,\ldots,q_n]$ be the set of orthonormal vectors,
such that
\begin{equation}\label{HQ_QT}
A^{\dag}A Q_n = Q_n T_n + \beta_n q_{n+1} (e_n^{(n)})^T,
~~~~q_1 = \rho_1 b, ~~~~\rho_1 = 1/||b||_2
\end{equation}
where $T_n$ is a tridiagonal and symmetric matrix,
$b$ is an arbitrary vector, and $\beta_n$ a real and positive constant.
$e_m^{(n)}$ denotes the unit vector with $n$ elements
in the direction $m$.

By writing down the above decomposition in terms of the vectors
$q_i, i=1,\ldots,n$ and the matrix elements of $T_n$, I arrive at
a three term recurrence that allows to compute these vectors
in increasing order, starting from the vector $q_1$. This is
the $Lanczos Algorithm$:
\begin{equation}
\begin{array}{l}
\beta_0 = 0, ~\rho_1 = 1 / ||b||_2, ~q_0 = o, ~q_1 = \rho_1 b \\
for ~i = 1, \ldots \\
~~~~v = A^{\dag}A q_i \\
~~~~\alpha_i = q_i^{\dag} v \\
~~~~v := v - q_i \alpha_i - q_{i-1} \beta_{i-1} \\
~~~~\beta_i = ||v||_2 \\
~~~~if \beta_i < tol, ~n = i, ~end ~for \\
~~~~q_{i+1} = v / \beta_i \\
\end{array}
\end{equation}
where $tol$ is a tolerance which serves as a stopping condition.

The Lanczos Algorithm constructs a basis for the Krylov subspace
\cite{Golub_VanLoan}:
\begin{equation}
\mbox{span}\{b,A^{\dag}Ab,\ldots,(A^{\dag}A)^{n-1}b\}
\end{equation}
If the Algorithm stops after $n$ steps, one says that the associated
Krylov subspace  is invariant.

In the floating point arithmetic, there is a danger that
once the Lanczos Algorithm (polynomial) has approximated well some part
of the spectrum,
the iteration reproduces vectors which are rich
in that direction \cite{Golub_VanLoan}.
As a consequence, the orthogonality of the Lanczos vectors is spoiled
with an immediate impact on the history of the iteration: if the
algorithm would stop after $n$ steps in exact arithmetic,
in the presence of round off errors the loss
of orthogonality would keep the algorithm going on.

\section{The Lanczos Algorithm for solving $A^{\dag}A x = b$}

Here I will use this algorithm to solve linear systems, where the loss
of orthogonality will not play a role in the sense that I will use
a different stopping condition.

I ask the solution in the form
\begin{equation}
x = Q_n y_n
\end{equation}
By projecting the original system on to the Krylov subspace I get:
\begin{equation}
Q_n^{\dag} A^{\dag}A x = Q_n^{\dag} b
\end{equation}
By construction, I have
\begin{equation}
b = Q_n e_1^{(n)} / \rho_1,
\end{equation}
Substituting $x = Q_n y_n$ and using (\ref{HQ_QT}), my task is
now to solve the system
\begin{equation}
T_n y_n = e_1^{(n)} / \rho_1
\end{equation}
Therefore the solution is given by
\begin{equation}
x = Q_n T_n^{-1} e_1^{(n)} / \rho_1
\end{equation}

This way using the Lanczos iteration one reduces the size of the
matrix to be inverted. Moreover, since $T_n$ is tridiagonal, one
can compute $y_n$ by short recurences.

If I define:
\begin{equation}
r_i = b - A^{\dag}A x_i, ~~q_i = \rho_i r_i, ~~\tilde{x}_i = \rho_i x_i
\end{equation}
where $i = 1, \ldots$, it is easy to show that
\begin{equation}
\begin{array}{l}
\rho_{i+1} \beta_i + \rho_i \alpha_i + \rho_{i-1} \beta_{i-1} = 0 \\
q_i + \tilde{x}_{i+1} \beta_i + \tilde{x}_i \alpha_i
+ \tilde{x}_{i-1} \beta_{i-1}  = 0
\end{array}
\end{equation}

Therefore the solution can be updated
recursively and I have the following
{\em Algorithm1 for solving the system $A^{\dag}A x = b$:}
\begin{equation}
\begin{array}{l}
\beta_0 = 0, ~\rho_1 = 1 / ||b||_2, ~q_0 = o, ~q_1 = \rho_1 b \\
for ~i = 1, \ldots \\
~~~~v = A^{\dag}A q_i \\
~~~~\alpha_i = q_i^{\dag} v \\
~~~~v := v - q_i \alpha_i - q_{i-1} \beta_{i-1} \\
~~~~\beta_i = ||v||_2 \\
~~~~q_{i+1} = v / \beta_i \\
~~~~\tilde{x}_{i+1} = - \frac{q_i + \tilde{x}_i \alpha_i
    + \tilde{x}_{i-1} \beta_{i-1}}{\beta_i} \\
~~~~\rho_{i+1} = - \frac{\rho_i \alpha_i + \rho_{i-1} \beta_{i-1}}{\beta_i} \\
~~~~r_{i+1} := q_{i+1} / \rho_{i+1} \\
~~~~x_{i+1} := y_{i+1} / \rho_{i+1} \\
~~~~if \frac{1}{|\rho_{i+1}|} < tol, ~n = i, ~end ~for \\
\end{array}
\end{equation}

\section{The Lanczos Algorithm for solving $(A^{\dag}A)^{1/2} x = b$}

Now I would like to compute $x = (A^{\dag}A)^{-1/2} b$ and still
use the Lanczos Algorithm. In order to do so I make the following
observations:

Let $(A^{\dag}A)^{-1/2}$ be expressed by a matrix-valued function,
for example the integral formula \cite{Higham}:
\begin{equation}
(A^{\dag}A)^{-1/2} = \frac{2}{\pi} \int_0^{\infty} dt (t^2 + A^{\dag}A)^{-1}
\end{equation}

From the previous section, I use the Lanczos Algorithm to compute
\begin{equation}
(A^{\dag}A)^{-1} b = Q_n T_n^{-1} e_1^{(n)} / \rho_1
\end{equation}

It is easy to show that the Lanczos Algorithm is shift-invariant.
i.e. if the matrix $A^{\dag}A$ is shifted by a constant say, $t^2$,
the Lanczos vectors remain invariant. Moreover, the corresponding
Lanczos matrix is shifted by the same amount.

This property allows one to solve the system
$(t^2 + A^{\dag}A) x = b$ by using the same Lanczos iteration
as before. Since the matrix $(t^2 + A^{\dag}A)$ is better conditioned
than $A^{\dag}A$, it can be concluded that once the
original system is solved, the shifted one is solved too.
Therefore I have:
\begin{equation}
(t^2 + A^{\dag}A)^{-1} b = Q_n (t^2 + T_n)^{-1} e_1^{(n)} / \rho_1
\end{equation}

Using the above integral formula and putting everything together,
I get:
\begin{equation}\label{result}
x = (A^{\dag}A)^{-1/2} b = Q_n T_n^{-1/2} e_1^{(n)} / \rho_1
\end{equation}

There are some remarks to be made here:

a) As before, 
by applying the Lanczos iteration on $A^{\dag}A$, the problem
of computing $(A^{\dag}A)^{-1/2} b$ reduces to the problem of computing
$y_n = T_n^{-1/2} e_1^{(n)} / \rho_1$ which is typically
a much smaller problem than the original one. But since $T_n^{1/2}$
is full, $y_n$ cannot be computed by short recurences.
It can be computed for example by using the full
decomposition of $T_n$ in its eigenvalues and eigenvectors; in fact
this is the method I have employed too, for its compactness and
the small overhead for moderate $n$.

b) The method is not optimal, as it would have been, if one would have
applied it directly for the matrix $(A^{\dag}A)^{1/2}$.
By using $A^{\dag}A$ the condition is
squared, and one looses a factor of two compared to the theoretical case!

c) From the derivation above, it can be concluded
that the system $(A^{\dag}A)^{1/2} x = b$
is solved at the same time as the system $A^{\dag}A x = b$.

d) To implement the result (\ref{result}),
I first construct the Lanczos matrix
and then compute
\begin{equation}
y_n = T_n^{-1/2} e_1^{(n)} / \rho_1
\end{equation}
To compute $x = Q_n y_n$, I repeat the Lanczos iteration.
I save the scalar products, though it is not necessary.

Therefore I have the following
{\em Algorithm2 for solving the system $(A^{\dag}A)^{1/2} x = b$:}
\begin{equation}
\begin{array}{l}
\beta_0 = 0, ~\rho_1 = 1 / ||b||_2, ~q_0 = o, ~q_1 = \rho_1 b \\
for ~i = 1, \ldots \\
~~~~v = A^{\dag}A q_i \\
~~~~\alpha_i = q_i^{\dag} v \\
~~~~v := v - q_i \alpha_i - q_{i-1} \beta_{i-1} \\
~~~~\beta_i = ||v||_2 \\
~~~~q_{i+1} = v / \beta_i \\
~~~~\rho_{i+1} = - \frac{\rho_i \alpha_i + \rho_{i-1} \beta_{i-1}}{\beta_i} \\
~~~~if \frac{1}{|\rho_{i+1}|} < tol, ~n = i, ~end ~for \\
\\
Set ~(T_n)_{i,i} = \alpha_i, ~(T_n)_{i+1,i} = (T_n)_{i,i+1} = \beta_i,
otherwise ~(T_n)_{i,j} = 0 \\
y_n = T_n^{-1/2} e_1^{(n)} / \rho_1
= U_n \Lambda_n^{-1/2} U_n^T e_1^{(n)} / \rho_1 \\
\\
q_0 = o, ~q_1 = \rho_1 b, ~x_0 = o \\
for ~i = 1, \ldots, n \\
~~~~x_i = x_{i-1} + q_i y_n^{(i)} \\
~~~~v = A^{\dag}A q_i \\
~~~~v := v - q_i \alpha_i - q_{i-1} \beta_{i-1} \\
~~~~q_{i+1} = v / \beta_i \\
\end{array}
\end{equation}
where by $o$ I denote a vector with zero entries
and $U_n, \Lambda_n$ the matrices of the eigenvectors and eigenvalues
of $T_n$. Note that
there are only four large vectors necessary to store: $q_{i-1},q_i,v,x_i$.

\section{Testing the method}

I propose a simple test: I solve the system $A^{\dag}A x = b$ by
applying twice the $Algorithm2$, i.e. I solve the linear systems
\begin{equation}
(A^{\dag}A)^{1/2} z = b, ~~(A^{\dag}A)^{1/2} x = z
\end{equation}
in the above order. For each approximation $x_i$, I compute the
residual vector
\begin{equation}
r_i = b - A^{\dag}A x_i
\end{equation}

The method is tested for a SU(3) configuration at $\beta = 6.0$
on a $8^316$ lattice, corresponding to an order $98304$ complex
matrix $A$.

In Fig.1 I show the norm of the residual vector decreasing monotonically.
The stagnation of $||r_i||_2$
for small values of $tol$ may come from the accumulation of round off
error in the $64$-bit precision arithmetic used here.

This example shows that the tolerance line is above the residual
norm line, which confirms the expectation that $tol$ is a good
stopping condition of the $Algorithm2$.

\section{Conclusions}

I have presented a Lanczos method to compute the inverse square root of
a large and sparse positive definite matrix.

The method is characterized by a residual vector norm
that decreases monotonically and a
consistent stopping condition. This stability
should be compared with a similar method presented earlier
by the author \cite{Borici}, where the underlying Hermitian but
indefinite matrix $\gamma_5 A$ led to appreciable instabilities
in the norm of the residual vector.

In terms of complexity this algorithm requires less operations
for the same accuracy than its indefinite matrix counterpart.
This property is guaranteed by the monotonicity of the residual vector
norm. Nontheless, the bulk of the work remains the same.

With the improvement in store the method is complete.

It shares with methods presented in \cite{Neuberger2,Edwards_Heller_Narayanan}
the same underlying Lanczos polynomial. As it is wellknown \cite{Golub_VanLoan}
CG and Lanczos methods for solving a linear system produce the same results
in exact arithmetic. In fact, CG derives from the Lanczos algorithm
by solving the coupled two-term recurences of CG for a single
three-term recurence of Lanczos.
However, the coupled two-term recurences of CG accumulate less round off.
This makes CG preferable for ill-conditioned problems.

There are two main differences between the method presented here and those in
\cite{Neuberger2,Edwards_Heller_Narayanan}:

a) Since CG and Lanczos are equivalent, they produce the same Lanczos matrix.
Therefore, any function of $A^{\dag}A$ translates for both algorithms into a
function of $T_n$ (given the basis of Lanczos vectors). The latter function
translates into a function of the Ritz values, the eigenvalues of $T_n$.
That is, whenever the
methods of papers \cite{Neuberger2,Edwards_Heller_Narayanan} try to approximate
the inverse square root of $A^{\dag}A$, the underlying CG algorithm shifts
this function to the Ritz values. It is clear now that if I take the inverse square root
from the Ritz values exactly, I don't have any approximation error. This is done
in $Algorithm2$.

b) $Algorithm2$ sets no limits in the amount of memory required,
whereas the multi-shift CG needs to store as many vectors as the number of shifts.
For high accuracy approximations the multi-shift CG is not practical. However, one may lift
this limit in expense of a second CG iteration (two-step CG) \cite{Neuberger3}.
Therefore $Algorithm2$ and the two-step CG have the same iteration
workload, with $Algorithm2$ computing exactly the inverse square root.

Additionally, $Algorithm2$  requires the calculation of Ritz eigenpairs of $T_n$,
which makes for an overhead proportional to $\sim n^2$ when using the QR algorithm
for the eigenvalues and the inverse iteration for the eigenvectors \cite{Golub_VanLoan}.
Since the complexity of the Lanczos algorithm is $\sim nN$, the
relative overhead is proportional to $\sim n/N$.
For moderate gauge couplings and lattice sizes this is a small percentage.

I conclude that the algorithms of \cite{Neuberger2,Edwards_Heller_Narayanan} may be
used in situations where a high accuracy is not required and/or $A$ is well-conditioned.

Experience with overlap fermions shows that high accuracy is often essential
\cite{Hern_Jans_Luesch,Borici}. In such situations the $Algorithm2$ is best suited.

\pagebreak

\pagebreak

\begin{figure}
\epsfxsize=12cm
\epsfxsize=10cm
\vspace{3cm}
\centerline{\epsffile[100 200 500 450]{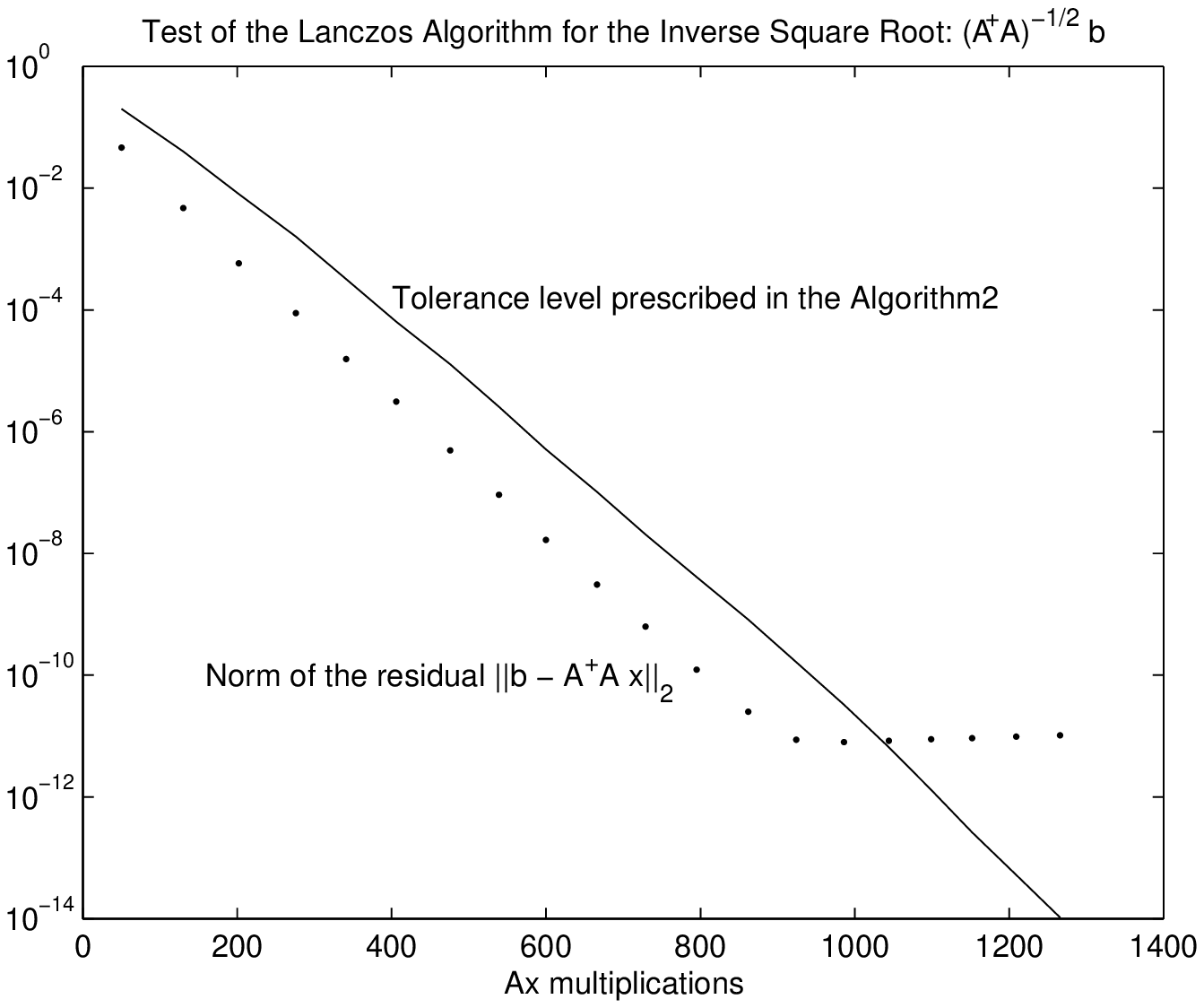}}
\caption{The dots show the norm of the residual vector, whereas
the line shows the tolerance level set by $tol$ in the $Algorithm2$.}
\end{figure}

\end{document}